\RequirePackage{fix-cm}
\documentclass[smallextended]{svjour3}       % onecolumn (second format)
\smartqed  % flush right qed marks, e.g. at end of proof
\usepackage{graphicx}
 \journalname{General Relativity and Gravitation.}
\begin{document}

\title{Accelerated Expansion of the universe based on Particle Creation-Destruction Processes and Dark Energy in FLRW universes.}


\author{Alberto C. Balfagon.
}


\institute{Seccion de Fisica Teorica y Aplicada. 
\\Departamento de Ingenieria Quimica,
\\Instituto Quimico de Sarria
\\Via Augusta, 390, 08017,Barcelona (Spain)
              Tel.: +34 932 67 20 00\\
              \email{albert.balfagon@iqs.url.edu}  }

\date{Received: date / Accepted: date}

\maketitle

\begin{abstract}
Particle creation has been considered as a possible justification for the accelerated expansion of the universe, obeying the second law of thermodynamics, together with the possible existence of Dark Energy. This paper introduces the possibility that the destruction of baryonic and/or dark matter particles also verifies the second law of thermodynamics thanks to a particle exchange with dark energy. General equations for the variation of the number of particles in accelerated universes have been obtained. Finally, a new model of the universe has been developed which predicts dark energy properties as well as particle exchange processes between dark energy and baryonic and/or dark matter.
\end{abstract}

\section{Introduction}
\label{sec:intro}

It seems well proven that the universe is experiencing an acceleration process [1-3]. Physicists have elaborated various answers [4] to explain this accelerated state, among them, the following may be mentioned: the existence of a new kind of energy (called 'dark energy' that  we could consider to include Einstein's famous cosmological constant) [5], matter creation [6-9], and a whole set of theories that modify or go beyond the General Theory of Relativity, such as F(R) [10].

The problem in elucidating what is really happening resides in the current data and the precision with which they are obtained. Many theories (and within them a whole set of different models) fit the available data, thus causing a high degree of degeneration in solving the question of the accelerating universe.

One of the ways of restricting this high degree of degeneration is by laying down the conditions (theoretical and experimental) the different models should fulfill. In the research we are presenting, we have obtained the general equation that - from a thermodynamically point of view [11,12] - the universe should fulfill in adiabatic conditions, when treated as an open system with particle creation or destruction [13-22]. From the previous equation, we have deduced the general equation a FLRW universe (treated as an adiabatic perfect fluid with particle creation or destruction) [23, 24] must verify by introducing the necessary terms in the energy-momentum tensor considering, as well, the possible existence of Dark Energy. In the present research, at no point the condition of constant specific entropy has been imposed [6, 24-26], only the condition of adiabatic open systems. 

Once the general condition an adiabatic FLRW universe with particle creation has been established and in order to delimit the solutions of the FLRW general equations, we have deduced the general restriction these universes must satisfy in order to show an accelerated expansion such as the one our universe is currently experiencing.

The rest of the article studies firstly and briefly the thermodynamics of open adiabatic systems with particle creation, then expanded Friedmann equations for FLRW universes are established, and the thermodynamic restriction that allows Friedmann equations to adapt to open systems with particle creation is set. In the next section, the general restriction an accelerated adiabatic universe with particle creation must satisfy is obtained. Later, two particular cases are analyzed: Prigogine model, and a new model developed for the first time (as far as the author is aware) in this article. To finish, the final conclusions are established.

\section{Thermodynamics of adiabatic open systems with matter creation.}

The first law of thermodynamics which applies to open systems (also called Gibbs equation for open systems) can be obtained from the first law of thermodynamics applicable to closed systems (Eq.1)

\begin{equation}
	dE = dQ - pdV
\end{equation}

Defining the energy density $(\rho)$ and the density of a particle (n)as  

\begin{equation}
	\rho= \frac {E}{V} ; n = \frac {N}{V}
\end{equation}

Where N and V are respectively the number of particles and the volume of the system.
Applying equation 2 on equation 1, considering N constant, and defining $ dq=dQ/N $ it becomes

\begin{equation}
	d(\frac {\rho}{n}) = dq - pd(\frac{1}{n})
\end{equation}

Equation 3 can be applied to open systems, if it is considered that N may not be constant.
Defining h as the enthalpy density, and supposing that N may not be constant, Gibbs equation for open systems is obtained from equation 3 (Eq.4)

\begin{equation}
	d(\rho V)= dQ -pdV + \frac {h}{n} d(nV)
\end{equation}

Considering adiabatic systems $(dQ = 0)$ equation 4 becomes (Eq.5)

\begin{equation}
	d(\rho V) + p dV - \frac {h}{n} d(nV) = 0
\end{equation}

Expanding equation 5 and dividing everything by $ Vdt $, it becomes (Eq.6)

\begin{equation}
	\dot \rho=\frac{h}{n} \dot n
\end{equation}

Note that in order to deduce equation 6 it has not been necessary to impose the condition of a constant specific entropy [6, 24-26], just the condition that the system is adiabatic (although, as it will be shown later, this condition implies that the specific entropy is constant).

Equation 6 can be applied to an adiabatic universe model where there is a variation in the number of particles either by their creation or by their elimination.

In order to establish the condition that must be fulfilled by an adiabatic FLRW universe with particle creation, it is preferable to express equation 6 in a different way. 

Defining a covolume $V=ka^3$, where k is a constant and a is the scale factor of the FLRW metric, it is easy to obtain (Eq.7)

\begin{equation}
	\frac {\dot n}{n} = \frac {\dot N}{N} -3H 
\end{equation}

Where H is the Hubbel parameter. Substituting equation 7 into equation 6, one gets (Eq.8)

\begin{equation}
	\dot \rho +3H(\rho + p)- \frac {\dot N}{N} (\rho+p) = 0
\end{equation}

The study of the entropy variation of the universe provides very useful information. In principle, entropy variation can come from two sources: the heat variation of the system, or the variation in its number of particles. Expressing this in the form of an equation, and using the definition of entropy, we have (Eq.9)

\begin{equation}
	TdS = dQ + T \frac{S}{N} dN
\end{equation}

Where T is the system temperature and the quotient $\frac{S}{N}$ is the specific entropy  $\sigma$.

Assuming that the universe is adiabatic, equation 9 is reduced to (Eq.10)

\begin{equation}
	dS = \sigma dN
\end{equation}

From equation 10, it follows that:

\begin{enumerate}
	\item The elimination of particles in the universe is not possible if the second law of thermodynamics is to be met.
	
	Dividing equation 10 by $dt$, one obtains $\dot S =\sigma \dot N$. According to the second law of thermodynamics $\dot S>0$, and since $\sigma$ is a positive quantity, consequently $\dot N >0$.
	\item 	Specific entropy is constant.
	
Calculating the total derivative of the expression that defines specific entropy, $S=N \sigma$, one obtains $dS=Nd\sigma + \sigma dN$, and comparing this expression with equation 10, it follows that  $d\sigma=0$, therefore, $\sigma$ is constant.
\end{enumerate}

\section{Expanded FLRW equations, thermodynamic conditions for the creation of matter.}

The universe is assumed to be an ideal fluid in which metric is given by the metric of the FLRW universe (Eq. 11).

\begin{equation}
 d s^2 = -c^2 d t^2 + a^2 ( \frac {d r^2} {1-\epsilon r^2} + r^2 d \theta^2 + r^2 Sin^2\theta d \Phi^2 )
\end{equation}

In order to obtain more general Friedmann equations, two terms are added to the energy momentum tensor $T_{\mu \nu}$. A first term (called A) is added to the energy term $T_{00}$, and a second term (called B) is added to the pressure terms $T_{ii}$. After carrying out the usual operations, a new set of Friedmann equations is deduced (Eq.12)

		\[H^2= \frac {8\pi G}{3c^2} \rho + \frac {c^2}{3} \Lambda -  \frac {c^2}{a^2} \epsilon + A
\]
	
\begin{equation}
  2\dot H +3H^2= - \frac {8\pi G}{c^2} p - \frac {c^2}{a^2}\epsilon + \Lambda c^2-B 
\end{equation}

Where c is the speed of light, G is the universal gravitation constant, and $\Lambda$ belongs to the possible term attributed to dark energy.

From the set of equations 12 it is easy to obtain the equation that allows us to deduce the condition a universe with matter creation must meet (Eq.13)

\begin{equation}
   \dot \rho +3H(\rho+p) + \frac {3c^2}{8\pi G}  [H(3A+B)+ \dot A + \frac {c^2 \dot \Lambda}{3}]=0 
\end{equation}

In the deduction of equation 13 , we have considered the general case where the terms A, B, and $\Lambda$  may vary with time.

By comparing equation 13 with its corresponding equation according to the thermodynamic development (Eq.8), we obtain the general condition that terms A and B must satisfy in order to have a universe with matter creation (Eq.14)

\begin{equation}
    - \frac {\dot N}{N} (\rho+p)=  \frac {3c^2}{8\pi G} [H(3A+B)+ \dot A + \frac {c^2 \dot \Lambda}{3}]
\end{equation}

If dark energy density is considered to be constant, or if its existence is simply not considered at all, equation 14 becomes (Eq.15)

\begin{equation}
   - \frac {\dot N}{N} (\rho+p)=  \frac {3c^2}{8\pi G} [H(3A+B)+ \dot A]
\end{equation}

According to equation 8, the variation of density with time is

\begin{equation}
   \dot \rho = - 3 H (\rho + p)+  \frac {\dot N}{N}  (\rho+p)     
\end{equation}

or, using equation 14

\begin{equation}
   \dot \rho =- 3 H (\rho+p)- \frac {3c^2}{8\pi G} [H(3A+B)+ \dot A + \frac {c^2 \dot \Lambda}{3}]  
\end{equation}

From equation 16, defining the quotient $\frac {\dot N}{N}$ as $\Gamma$, and using the definition of enthalpy density h, we get

\begin{equation}
   \dot \rho = h(\Gamma-3H)
\end{equation}

Taking into account that $\frac {\dot n}{n} = \Gamma-3H$, equation 18 is finally expressed as equation 19, which matches the one obtained by other authors [23]

\begin{equation}
   \dot \rho = h \frac {\dot n}{n}
\end{equation}

\section {Accelerated expansion of the universe.}
Since the universe is currently experiencing an expansion phase, it is advisable to lay down the condition for the existence of this acceleration, which will also allow for further delimiting the possible cosmological models.

This condition has been obtained starting from the calculation of the expression given by equation 20
\begin{equation}
   2 \frac {\ddot a}{a}>0     
\end{equation}

Whose sign will depend only on the second time derivative of the scale factor, since a is always positive.

Starting from Friedmann equations (Eq.12), and using equations 8 and 14, the general condition the accelerated universe must meet is obtained (Eq.21)

\begin{equation}
   -\frac {8\pi G}{3c^2} (\rho+3p)+  \frac {2c^2}{3} \Lambda - (A+B)  >0     
\end{equation}

Eliminating the dark energy term, the final expression becomes (Eq.22)
\begin{equation}
   -\frac {8\pi G}{3c^2} (\rho+3p)- (A+B)  >0 
\end{equation}

It follows that the transition points from an accelerated phase to a decelerated one must verify (Eq.23)
\begin{equation}
    -\frac {8\pi G}{3c^2} (\rho+3p)= (A+B)     
\end{equation}

\section{Particular cases.}

In this section two specific cases of singular importance are studied starting from previously developed equations.

\subsection{Prigogine model.}

This model studied by Prigogine [24] and other authors [21, 23], assumes that parameters A and $\Lambda$ are null, so equations 14 and 21 become

\begin{equation}
   -\frac {\dot N}{N} (\rho+p)=  \frac {3c^2}{8\pi G} HB
\end{equation}
\begin{equation}
   -\frac {8\pi G}{3c^2} (\rho+3p)- B >0     
\end{equation}

Since H has the same dimensions as the quotient $\frac {\dot N}{N}$, it follows from equation 24 that the term B can be interpreted as a pressure multiplied by some suitable dimensional constant. The constant is easily deduced from Friedmann's second equation [12]

\begin{equation}
   B= \frac {8\pi G}{c^2}  p_c     
\end{equation}

Where $p_c$ is the creation pressure caused by the creation of particles. After solving equation 24 for B, and comparing it with equation 26, we have
\begin{equation}
   p_c= -\frac {h}{3H} \frac {\dot N}{N}     
\end{equation}

From equation 27 follows that matter creation creates a negative pressure. After some simple transformations, equation 27 can be expressed in the same form as the one found in [23] (equation 28) or in the form given by Prigogine [24] (equation 29)

\begin{equation}
   p_c= -\frac {h}{3H} (3H+\frac {dLn(n)}{dt})     
\end{equation}

\begin{equation}
   p_c= -\frac {h}{n} \frac {d(nV)}{dV}
\end{equation}

The condition that particle creation velocity must meet in so that the universe experiments an accelerated expansion is obtained by combining equations 25, 26, and 27 (equation 30)

\begin{equation}
   \dot N  > \frac {(\rho+3p)NH}{h}
\end{equation}

If we assume that pressure follows equation of state of the $p=W\rho$ type, equation 30 becomes equation 31 which coincides with the one deduced in [21] when $W = 0$

\begin{equation}
   \dot N  > \frac {1+3W}{1+W}  HN     
\end{equation}

\subsection{Transmutation model T}

This model, described for the first time in this article (as far as the author knows), is defined by the following parameter values: $A = B = 0, \Lambda \neq 0$.

With the aforementioned values, equations 14 and 21 are now expressed as

\begin{equation}
   -\frac {\dot N}{N} (\rho+p)=  \frac {3c^2}{8\pi G} \frac {c^2 \dot \Lambda}{3}   
\end{equation}

\begin{equation}
   \frac {8\pi G}{3c^2} (\rho+3p)<  \frac {2c^2}{3} \Lambda
\end{equation}

Before analyzing the different submodels that can be deduced from equations 32 and 33, it is necessary to analyze again equations 8, 10, 12 and 13 in the case dark energy exists.

The first Equation 12 shows that when $A=B=0$ the energy content of the universe is divided in three large groups: the group defined by the density $\rho$ attributable to baryonic and dark matter, the curvature energy (which depends on the value of $\epsilon$), and finally the dark energy variable (which depends on the value of $\Lambda$).

Equation 8 can be applied only to baryonic and to dark matter as reflected when comparing equations 8 and 13 in order to obtain equation 14. From what has been stated before, it can be observed that in equation 32 the number of particles N and its temporal variation $\dot N$ relate only to baryonic and dark matter. It should be also noted that if the pressure p corresponds to baryonic and dark matter then it follows that the pressure of dark energy would be zero (which is consistent with the interpretation to be given further on about the possible conversion between dark energy and baryonic or dark matter).

Regarding the entropy of the universe (equation 10), there are two possible interpretations:

\begin{enumerate}
	\item Either the entropy of the universe depends only on baryonic and dark matter, or the specific entropy of dark energy is zero. In both cases equation 10 is obtained, therefore, according to the second law of thermodynamics $\dot S>0$  which implies that (Eq.34)
	\begin{equation}
	   \dot N>0
	\end{equation}
	\item The entropy of the universe is a function of all its components, therefore, equation 10 becomes (Eq.35)
	\begin{equation}
	   dS=\sigma dN+\sigma_\Lambda dN_\Lambda
	\end{equation}        
	Where $\sigma_\Lambda$ is the specific entropy of dark energy and $N_\Lambda$ is the number of dark energy particles.
	
	From the second law of thermodynamics (taking into account that the condition $Nd\sigma+N_\Lambda d\sigma_\Lambda=0$ is verified, as deduced from $S=N\sigma+N_\Lambda \sigma_\Lambda$ and that the universe is an adiabatic system), it follows that (Eq.36)
	\begin{equation}
	   \dot S = \sigma \dot N +\sigma_\Lambda \dot N_\Lambda >0     
	\end{equation}
\end{enumerate}

From equations 32 and 33 different submodels can be deduced, which can be classified in two main groups

\subsubsection{T-I Model}
Particles are created $(\dot N >0)$ and dark energy density decreases with time. This decrease of dark energy can be interpreted as a consequence of the expansion of the universe, or as the sum of two processes: the decrease caused by the expansion of the universe plus a destruction of dark energy that could trigger the creation of baryonic and/or dark matter.

The entropy of the universe would increase over time if one assumes the case of equation 34, or if one assumes the case given by equation 36 as long as the condition established by this equation is met.

By operating in the same way with the Prigogine model, the acceleration would be caused according to equation 32 by a negative pressure due to the creation of particles.

A possible interpretation of the evolution of the universe - a qualitative one - is shown in figure 1, where the Y-axis represents both terms of equation 33, and the horizontal axis represents time. In this interpretation there is some dark energy in the universe prior to the inflationary and nucleosynthesis period. Later, the universe would continue expanding in a decelerated way (all densities would decrease - at different speeds - due to the expansion of the universe even though there could be particle creation due to the mechanism mentioned above). From the intersection point of both curves , the universe would enter again in an acceleration phase, despite the fact that dark energy density would continue to decrease. The subsequent evolution of the universe would lead to the elimination of dark energy (in the case where its particles are being destroyed), ending the production of baryonic and/or dark matter particles.

\begin{figure}
\centering
	\includegraphics[width=0.7\textwidth]{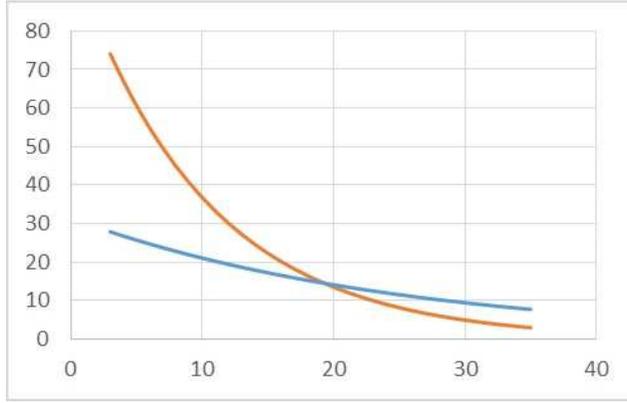}
\caption{Qualitative representation of both terms of equation 33 (separately, on the Y-axis) vs. time (X-axis), arbitrary units in both axes. The almost linear curve represents the dark energy component. The intersection point correspond to change in the sign of the acceleration of the universe.}
\label{fig:1}       
\end{figure}

\subsubsection{T-II Model}
In this model, particles are destroyed $(\dot N<0)$ and the dark energy density increases with time. In this case two situations may arise:

\begin{enumerate}
	\item If either the entropy of the universe is only a function of baryonic and dark matter, or the specific entropy or dark energy is zero, then this model is not possible if the second law of thermodynamics must be fulfilled. The other possibility is that this model would indicate that the second law of thermodynamics is not true and entropy may decrease with time.
	\item If equation 36 holds, this model could be compatible with the second law of thermodynamics. The decrease of baryonic and/or dark matter entropy would be compensated by an increase of dark energy entropy.
\end{enumerate}

In this model, the origin of the acceleration of the universe could be interpreted as an anti-gravitational effect or a weakening of the gravitational field cause by dark energy (this interpretation would relate to the ideas proposed by other authors [27-29]).

\begin{figure}
\centering
 \includegraphics[width=0.75\textwidth]{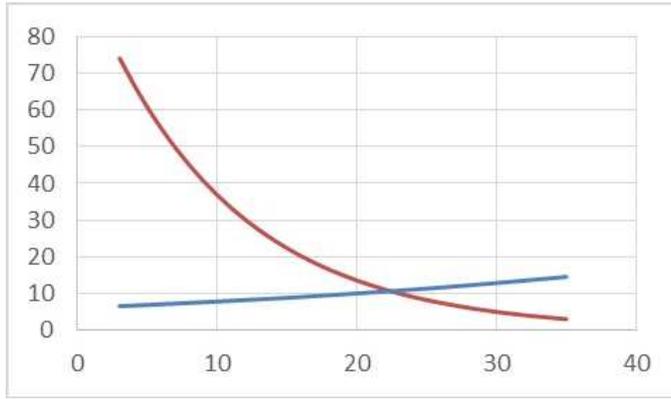}
\caption{Qualitative representation of both terms of equation 33 (separately, on the Y-axis) vs. time (X-axis), arbitrary units in both axes.  The almost linear curve represents the dark energy component. The intersection point correspond to change in the sign of the acceleration of the universe.}
\label{fig:2}       % Give a unique label
\end{figure}

A possible interpretation of the evolution of the universe - a qualitative one - is shown in figure 2, where the Y-axis represents both terms of equation 33, and the horizontal axis represents time. In this interpretation there is also some dark energy in the universe prior to the inflationary and nucleosynthesis period. Later, the universe would continue to expand in a decelerated way. Baryonic energy and dark matter density would decrease (due to the expansion and to the elimination of particles) while dark energy density would increase (despite the expansion of the universe, this density could increase considering that dark energy is the result of the sum of a constant vacuum energy and the energy caused by the destruction of particles). From the intersection point of both curves, the universe would enter again in an acceleration phase. The subsequent evolution of the universe would lead to the complete destruction of baryonic and/or dark matter.

\section{Conclusions}
In this research we have obtained a general equation (Eq.14) that relates the variation of the total number of baryonic or dark particles (in a flat, open or closed, adiabatic FLRW universe) to the components of the universe (either baryonic or dark matter or dark energy). This equation has been obtained by balancing  equation 8 - equation obtained from the thermodynamic principles of adiabatic open systems (and applied  to baryonic and dark matter and assuming that dark matter will comply with the same laws) - with the corresponding part of the equation 13 obtained from the expanded Friedmann equations.

The general condition (Eq.21) that must be met for the universe to show stages of accelerated expansion has been deduced, considering the possibility of a variation in the total number of particles.

From the second law of thermodynamics follows that if the entropy of the universe is a function only of the baryonic or dark matter, there must be particle creation. However, if dark energy may vary over time, then in order to fulfill the second law of thermodynamics the condition given by equation 36 must be satisfied. Also, there is a possibility of particle exchange between dark energy and baryonic and/or dark matter.

From the general equations 14 and 21, two particular cases have been studied. The first one is a model originally studied by Prigogine [24] and later by other authors [21, 23]. The second case, the Transmutation Model has been studied for the first time in this research (as far as the author knows).

From the Transmutation model follow two possible evolutions of the universe (in both cases fulfilling the second law of thermodynamics). According to the first one (the T-I Model) there would be a creation of baryonic and/or dark matter particles from dark energy. The cause of the accelerated process would be explained - in a similar way as the Prigogine model - by a negative pressure.

According to the second possibility (T-II model) there would be a destruction of baryonic and/or dark matter particles resulting in a generation of dark energy particles. In this case, the reason for the accelerated expansion of the universe should be sought in a weakening of the gravitational force caused by the presence of dark energy, or by the possibility that dark energy shows the phenomenon of anti-gravitation.

Taking into account the real lack of knowledge about what dark energy is (in case it exists), it is always interesting to deduce new models that predict any of its properties. The Transmutation model predicts a particle exchange between baryonic and/or dark matter and dark energy, as well as the possible existence of some anti-gravitational or gravity weakening effect. Particle physicists and astrophysicists may have some say in this matter.

\end{document}